\documentclass[prl,aps,twocolumn]{revtex4}
\usepackage{bm}
\usepackage{graphicx}
\usepackage{amsfonts}
\usepackage{amsmath}
\usepackage{amssymb}
  
\newcommand {\be}{\begin{equation}}
\newcommand {\ee}{\end{equation}}
\newcommand {\bea}{\begin{eqnarray}}
\newcommand {\eea}{\end{eqnarray}}

\begin{document}

\title{Optimal control of electromagnetic field using metallic nanoclusters}
\author{Ilya Grigorenko}
\affiliation{T-11, Center for Nonlinear Studies, Center for Integrated Nanotechnologies,
Los Alamos National Laboratory, Los Alamos, NM 87545}
\author{Stephan Haas}
\affiliation{Department of Physics and Astronomy, University of Southern California, Los
Angeles, CA 90089-0484}
\author{Alexander Balatsky}
\affiliation{T-11, Center for Integrated Nanotechnologies, Los Alamos National
Laboratory, Los Alamos, NM 87545}
\author{A.F.J. Levi}
\affiliation{Department of Physics and Astronomy, University of Southern California, Los
Angeles, CA 90089-0484}
\affiliation{Department of Electrical Engineering, University of Southern California, Los
Angeles, CA 90089-2533}
\date{\today}

\begin{abstract}
The dielectric properties of metallic nanoclusters in the presence
of an applied electromagnetic field are investigated using
non-local linear response theory. In the quantum limit we find a
non-trivial dependence of the induced field and charge
distribution on the spatial separation between the clusters and on
the frequency of the driving field. Using a genetic algorithm,
these quantum functionalities are exploited to custom-design
sub-wavelength lenses with a frequency controlled switching
capability.
\end{abstract}

\maketitle


\textit{Introduction:} Recently we developed a theory that describes
the non-local linear dielectric response of nano-metal structures to
an externally applied electric field \cite{levi}. Unlike
conventional phenomenological classical theory \cite{mie}, we are
able to model the transition from classical to quantum response
as well as the coexistence of classical and quantum response in structures
of
arbitrary geometry. This is of some practical importance because metallic
nanoclusters can now be made sufficiently small such that
non-local effects due to finite system size and cluster shape
dominate the spectral response. In particular, when the ratio between the
smallest
characteristic length scale and the Fermi wavelength is comparable to or
smaller than unity, these systems can fail to fully screen
external driving fields. Also, in the quantum limit one needs to
take into account discreteness of the excitation spectrum as well
as the intrinsically strong damping of collective modes. Our model
captures these single and many particle quantum effects.

\qquad In this paper we report on studies in which we explore
optimal design of nanoscale metallic structures to control
electromagnetic field intensity on subwavelength scales. We are
motivated by the non-intuitive nature of quantum response and the
potential for applications such as surface enhanced Raman scattering
\cite{sers,nurmikko}.

\textit{Model:} To capture the single-particle and collective aspects
of light-matter interaction in inhomogeneous nanoscale systems one
should consider non-local response theory in the quantum regime
\cite{levi}. The Schr\"odinger equation for noninteracting
electrons with mass $m_{\text e}$ and charge $e$ moving in a
potential $V(\mathbf{r})$, is given by
\begin{equation}  \label{hamiltonian}
H\Psi_i(\mathbf{r}) = \left( -\frac{\hbar^2}{2m} \nabla^2 + V(\mathbf{r})
\right) \Psi_i(\mathbf{r}) = E_i \Psi_i(\mathbf{r}).
\end{equation}
This equation must be solved simultaneously with the
Poisson equation to determine the local potential due to the
spatial distribution of the positive background charges. Using the
jellium approximation, the resulting potential is implicitly given
by $\nabla^2 V(\mathbf{r})=4 e \pi \rho(\mathbf{r})$, where the density
of the positive background charge $\rho(\mathbf{r})$ satisfies the
condition of neutrality, so that $\int \rho(\mathbf{r})d \mathbf{r}=N_{%
\text {el}}$, where $N_{\text {el}}$ is the number of electrons. The
induced potential $\phi _{\text{ind}}$ is then determined from the
self-consistent integral equation \bea \label{integral_equation}
\phi _{{\text {ind}}} ({\bf r},\omega) = \int{ {\frac{\chi ({\bf
r'},{\bf r''},\omega )\phi _{{\text {tot}}} ({\bf r''},\omega)
}{{\left| {{\bf r} - {\bf r'}} \right|}}d{ \bf r'}} d {\bf r''}}.
\eea Here $\chi({\bf r'},{\bf r''},\omega )$ is the
non-local density-density response function, and $\phi _{\text {tot}} (%
\mathbf{r},\omega)=\phi _{\text {ext}} (\mathbf{r},\omega) +
\phi _{\text {ind}} (\textbf{r},\omega)$ is the self-consistent total
potential. The induced field is found via $\mathbf{E}_{{\text {ind}}}(%
\mathbf{r},\omega)=-\mathbf{\nabla} \phi _{{\text {ind}}}
(\textbf{r},\omega)$. The external field is assumed to be harmonic,
with frequency $\omega$, linearly polarized, and with the
wavelength much larger than the characteristic system's size. The
integral equation is discretized on a real-space cubic mesh with
lattice constant $L$. Its natural energy scale $E_0$ is defined by
$E_0=\hbar^2/(2 m_{\text e} L^2)$, and the resulting system of
linear equations can be solved numerically. The damping constant
which determines the level broadening is set to $\gamma=2\times
10^{-3}$ $E_0$.

\begin{figure}[tbp] 
\includegraphics[width=2.81cm,angle=0]{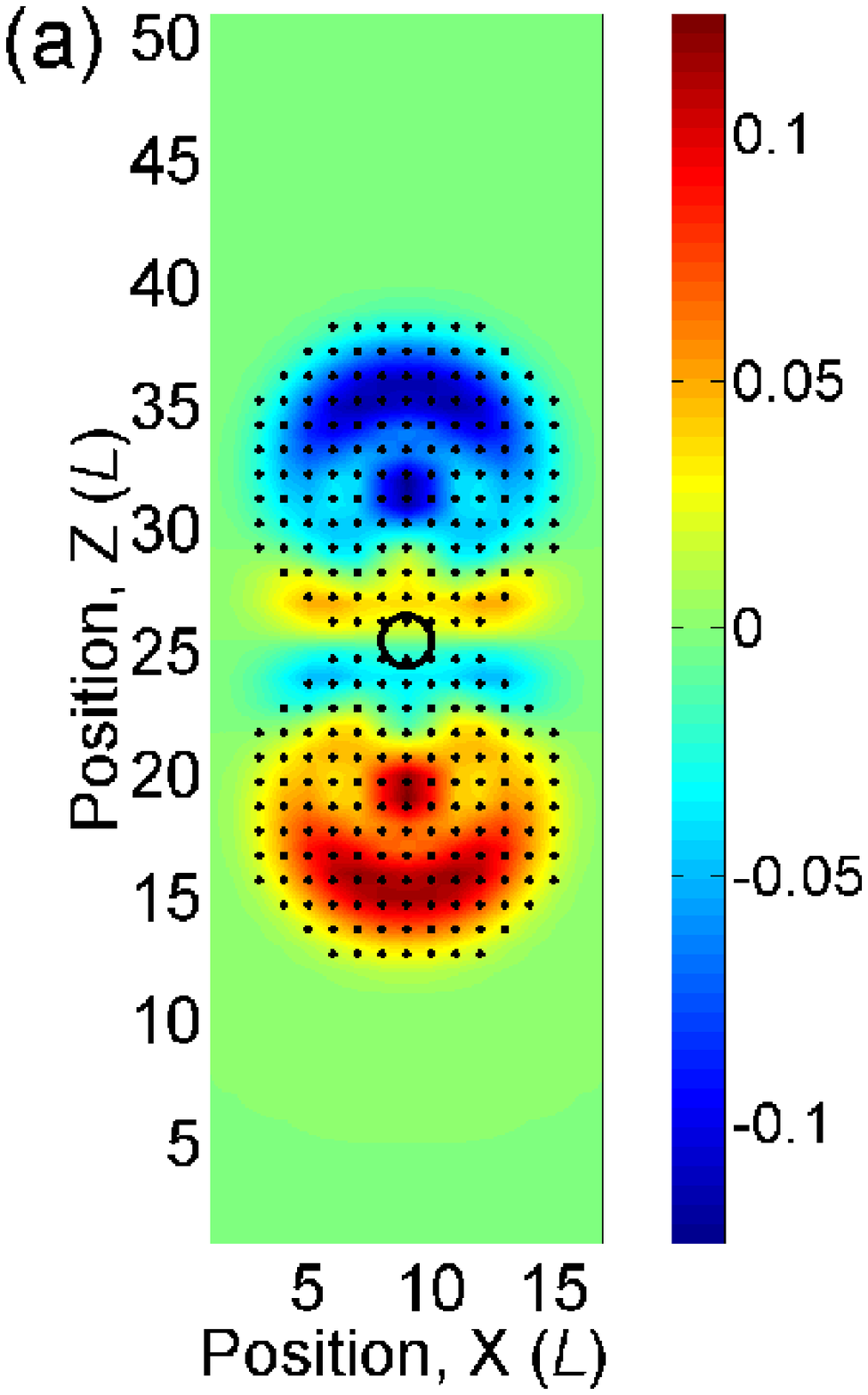}
\includegraphics[width=2.81cm,angle=0]{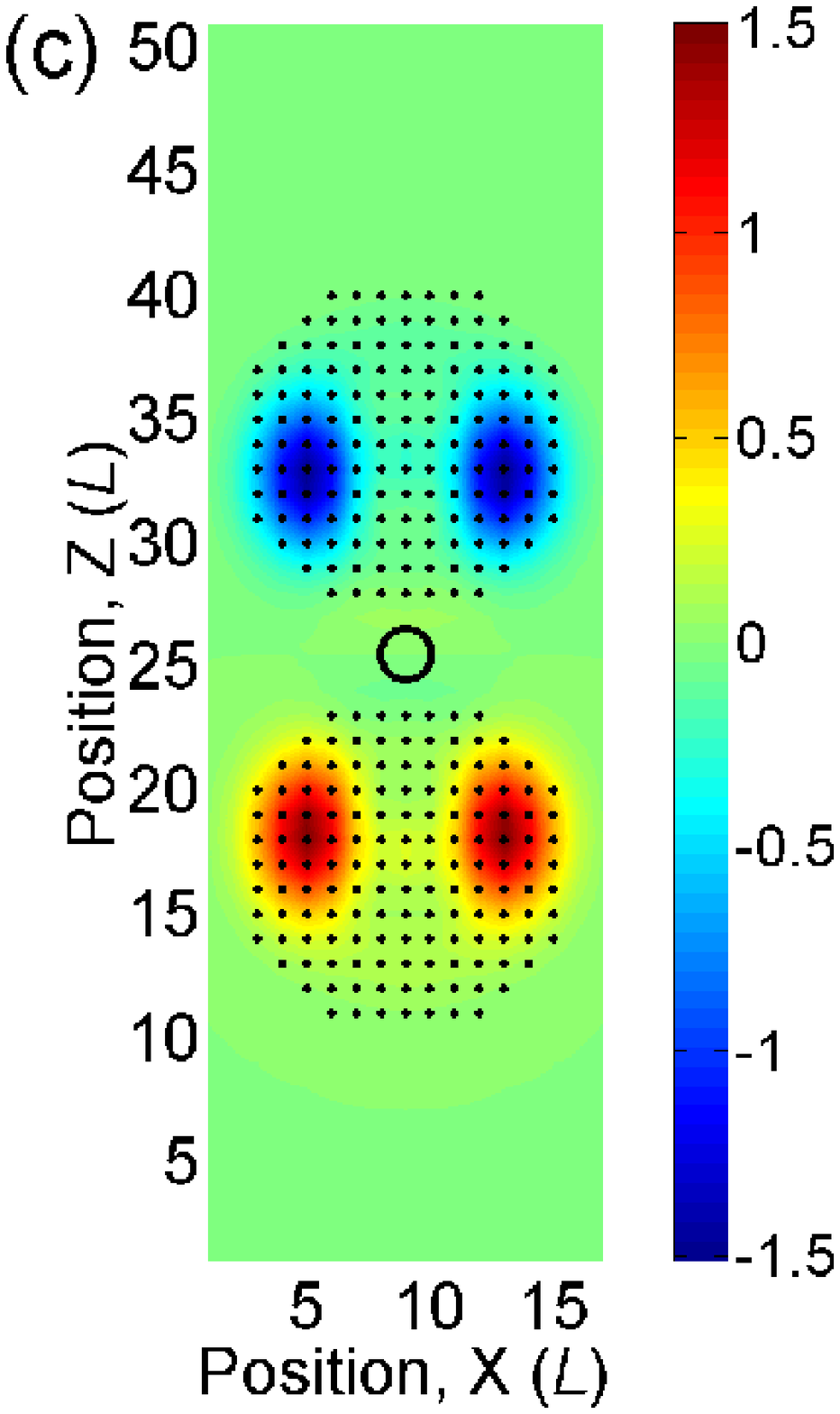}
\includegraphics[width=2.81cm,angle=0]{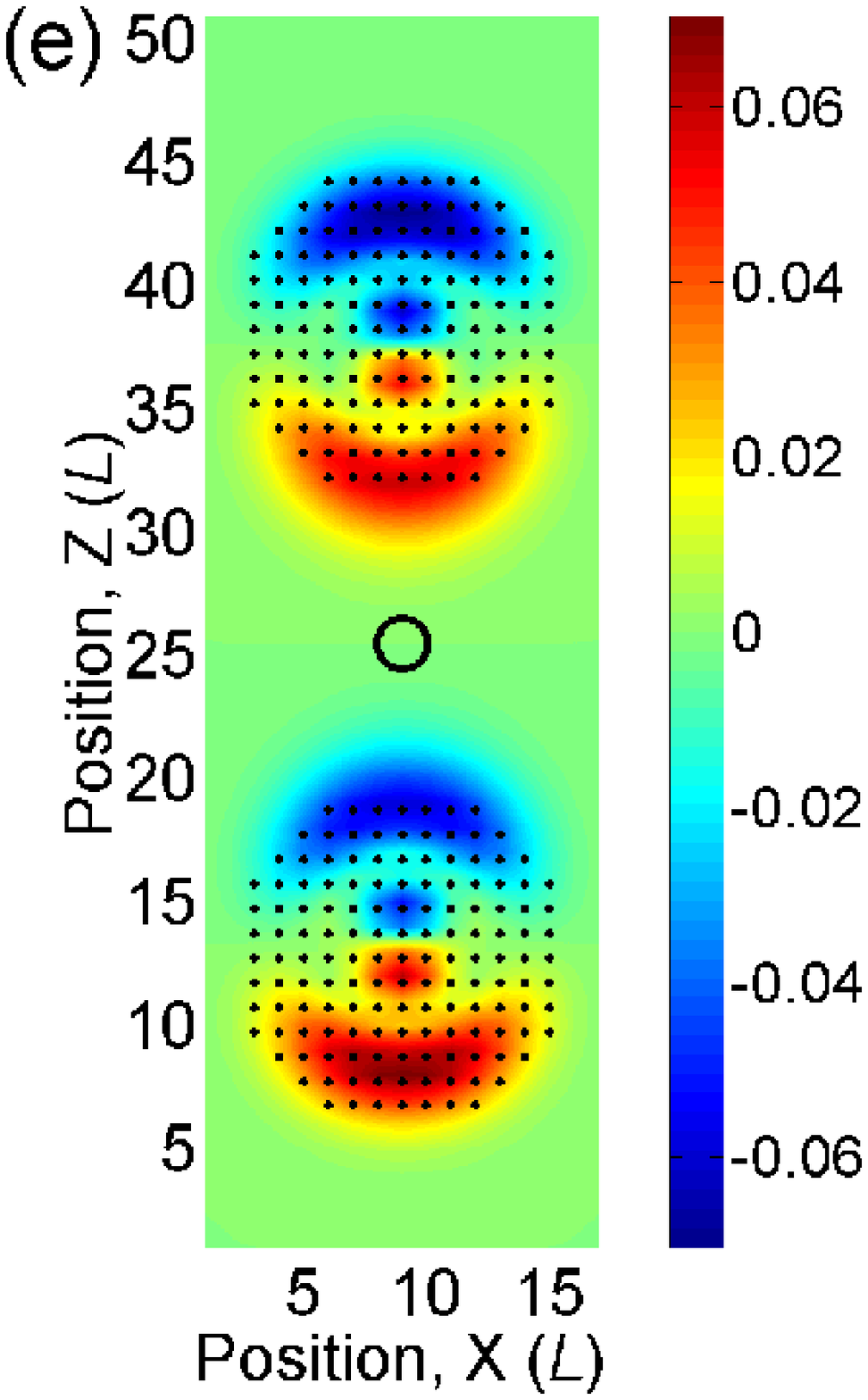}
\includegraphics[width=2.81cm,angle=0]{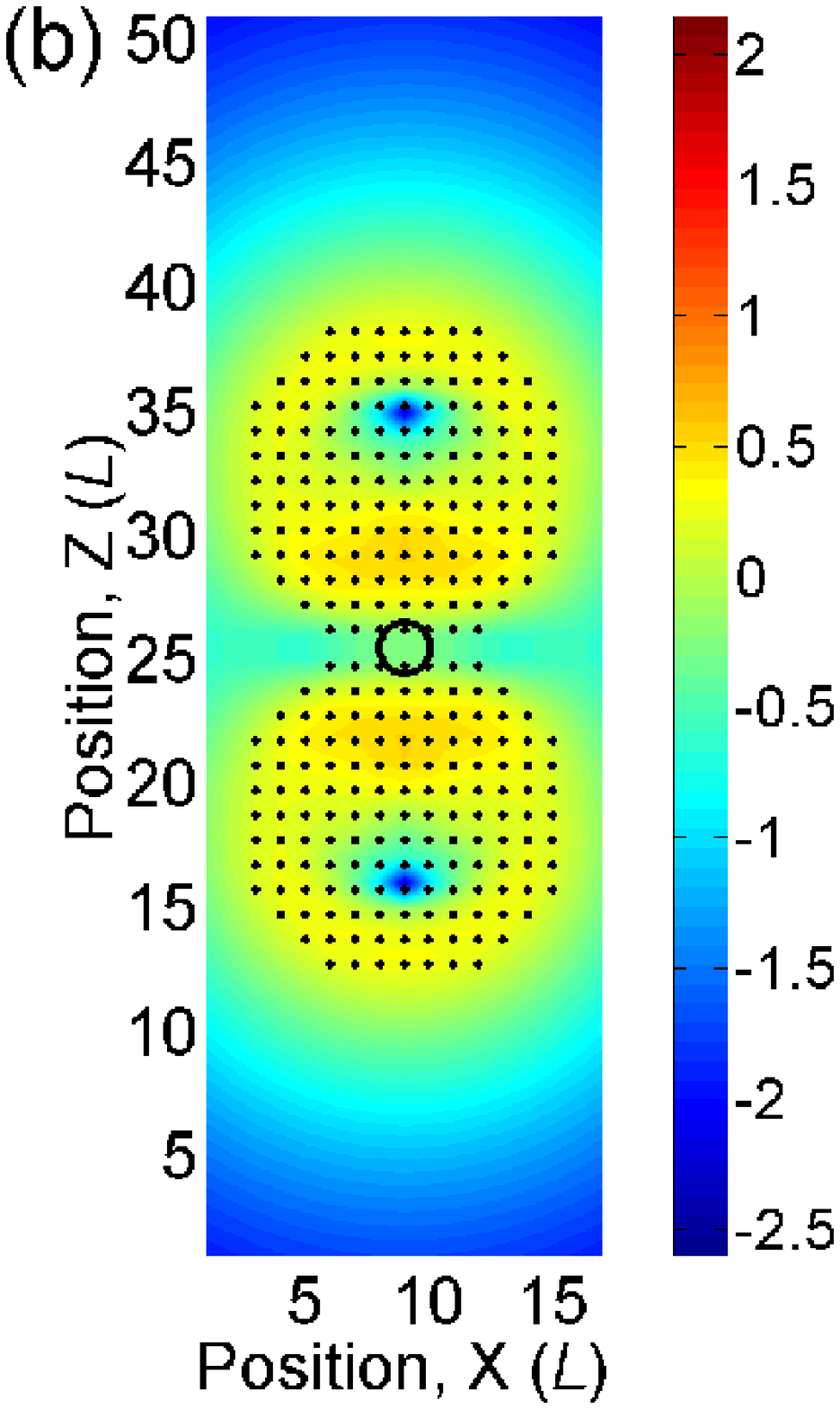}
\includegraphics[width=2.81cm,angle=0]{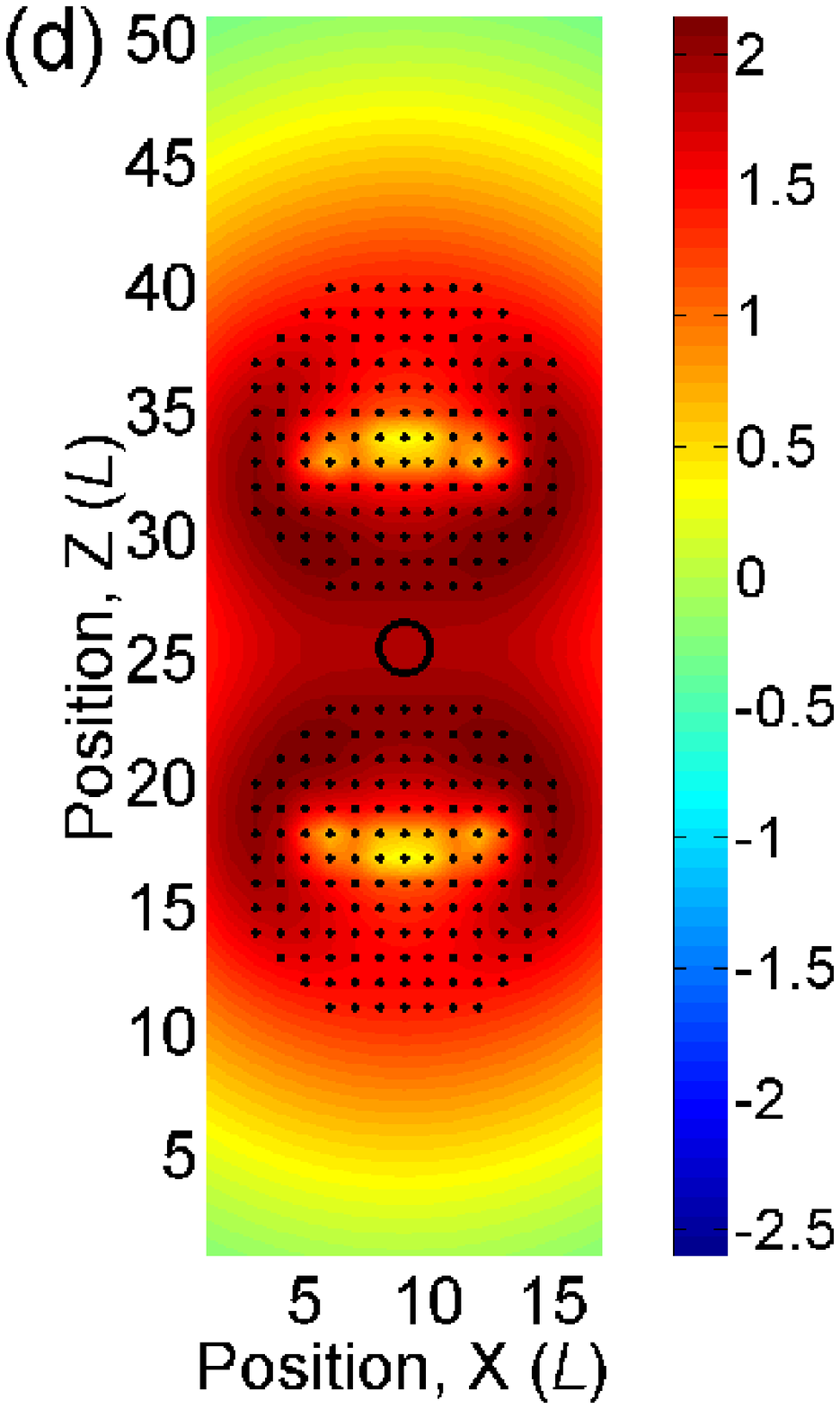}
\includegraphics[width=2.81cm,angle=0]{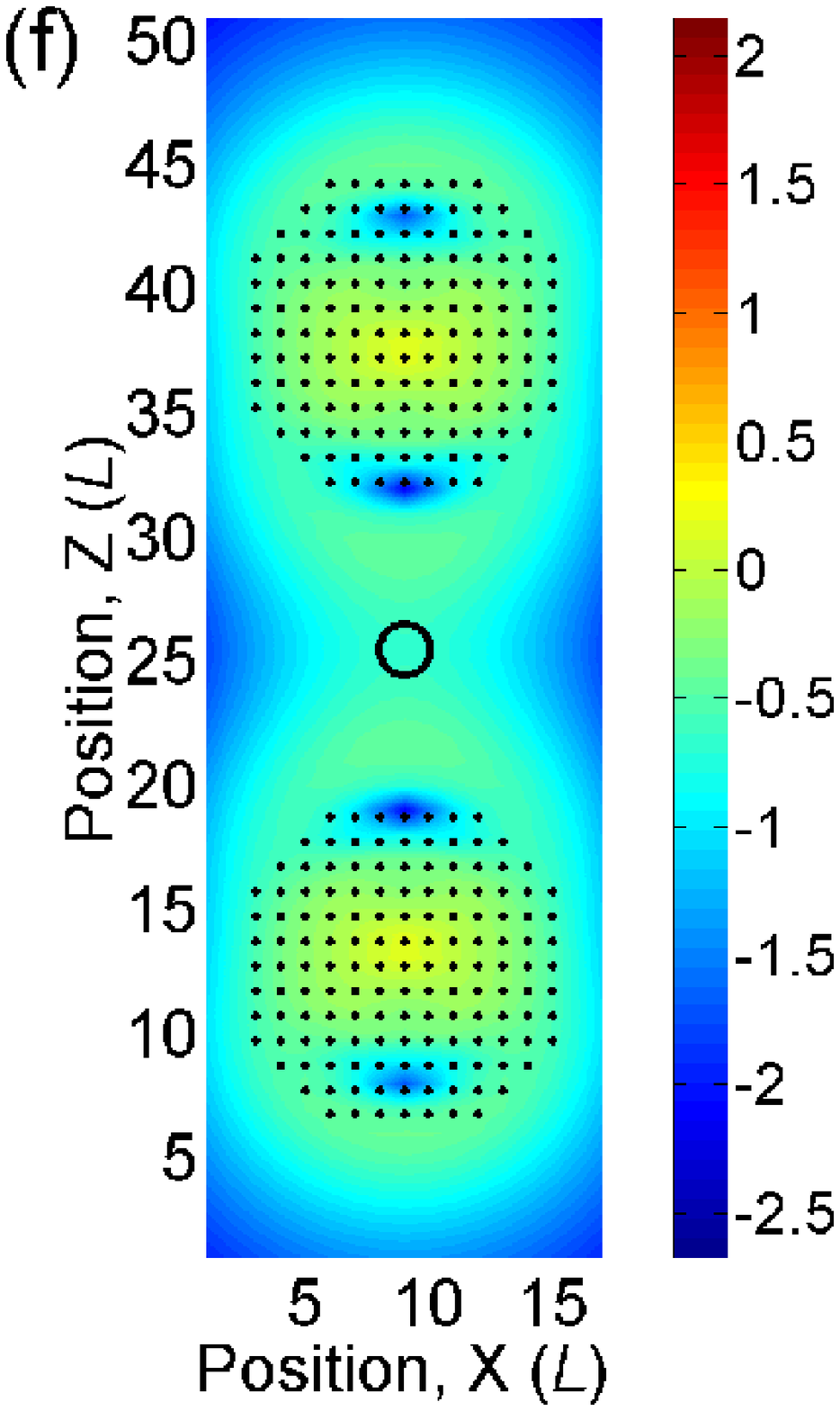}
\vspace{-0.5cm}
\caption{{\protect\footnotesize {\ The direction of the static external
field $\mathbf{E}_{\text{ext}}(\protect\omega = 0)$ points along the $z$
axis. Induced charge density (upper row) and corresponding induced electric
field $\log_{10}(|\mathbf{E}_{\text{ind}}(x,0,z)|^2)$ (lower row) in systems
of two conducting spherical clusters with radii $R=7 L$ (shown by black
dots), where $L$ is the lattice spacing. The plots show top views of the
three-dimensional systems. The relative distance $D$ between the two closest
points of the spheres is varied from left to right: (a),(b): $D$=0; (c),(d):
$D=4 L$, (e),(f): $D=13 L$. $N_{\text{el}}=20$ for the system, temperature $T$=0 K and damping
constant $\protect\gamma=2\times 10^{-3}$ $E_0$. The target volume over
which the field intensity is to be maximized is indicated as a circle of
radius $L$ at the system center. }}}
\label{fig1}
\end{figure}

\textit{Static Case:} To gain better understanding and intuition into the
dielectric response
of nanoscale clusters in the quantum limit, let us first consider
systems consisting of two identical nanospheres with a total
number of electrons $N_{\text {el}}$ and separated by an
adjustable distance $D$ (top view shown in Fig. 1). The
nanospheres are placed in a static electric field, and the
z-direction of the external field is aligned along the line
connecting the sphere centers. Our goal is to maximize the
intensity of the induced electric field
$W_{\text{int}}=\int_{V_0}|\mathbf{E}(\mathbf{r})|^2 d \mathbf{r}$ in a
target volume of radius $R_{\text V}= L$, centered between the two
clusters, by varying the cluster separation $D$. In the regime of
large electron densities, classical theory predicts that
$W_{\text{int}}$ diverges as the spheres approach each other, i.e.
$W_{\text{int}}\to\infty$ as $D\to 0$, and hence the spherical
clusters would need to be as close as possible to each other to
maximize the induced field in the target area. As seen in Fig. 1,
this is no longer true for small carrier concentrations (here
$N_{\text{el}}=20$), in which case quantum fluctuations strongly
influence the electromagnetic response. For sufficiently small
separations (Fig. 1(a) and (b)) the entire system responds as a
single dipole (with small corrections at the interface between the
two clusters). The charge density distribution depicted in Fig.
1(a) shows charge polarization (red: positive, blue: negative)
along the applied field, whereas the corresponding induced
electric field in Fig. 1(b) remains relatively homogeneous
throughout the entire system. Remarkably, as shown in Fig. 1(c)
and (d), there are \textit{finite optimum separations} between the
spheres which maximize the induced field at the center between
them. As will become apparent, the physical reason for this is
quantum wave functions constrain accessibility to geometric features.
For the parameters chosen in this example, $D_{\mathrm{opt}}\approx 4L$
occurs at the threshold separation distance
beyond which the two clusters cease to act as a single dipole. It
is evident from Fig. 1(d) that at this resonance condition the
overall induced field intensity is highly inhomogeneous and peaks
at two orders of magnitude larger compared to off-resonance
conditions. Moreover, there can be further such resonances, e.g.
at $D_{\mathrm{opt}}\approx 7L$ for the present parameters, which
maximize the induced field intensity in between the nanospheres.
As observed in Fig. 1(e) and (f), for larger distances $D$ one
can ultimately treat the spheres as independent dipoles, for which
the induced field energy $W_{\text{int}}$ scales as
$D^{\lambda}$, with $\lambda=-8$, which we have verified
numerically \cite{footnote1}.

\begin{figure}[tbp]
\includegraphics[width=4.25cm,angle=0]{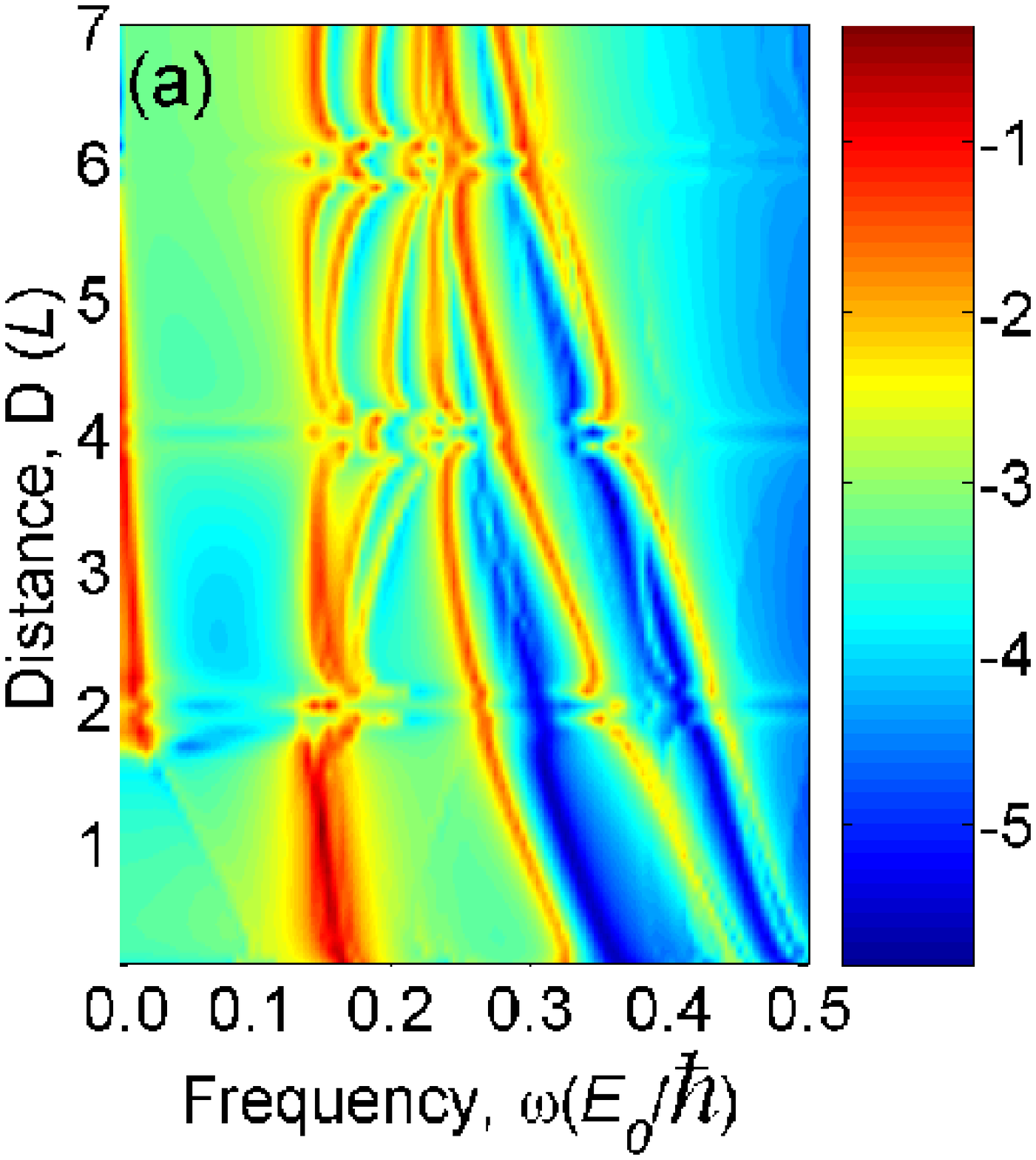}
\includegraphics[width=4.25cm]{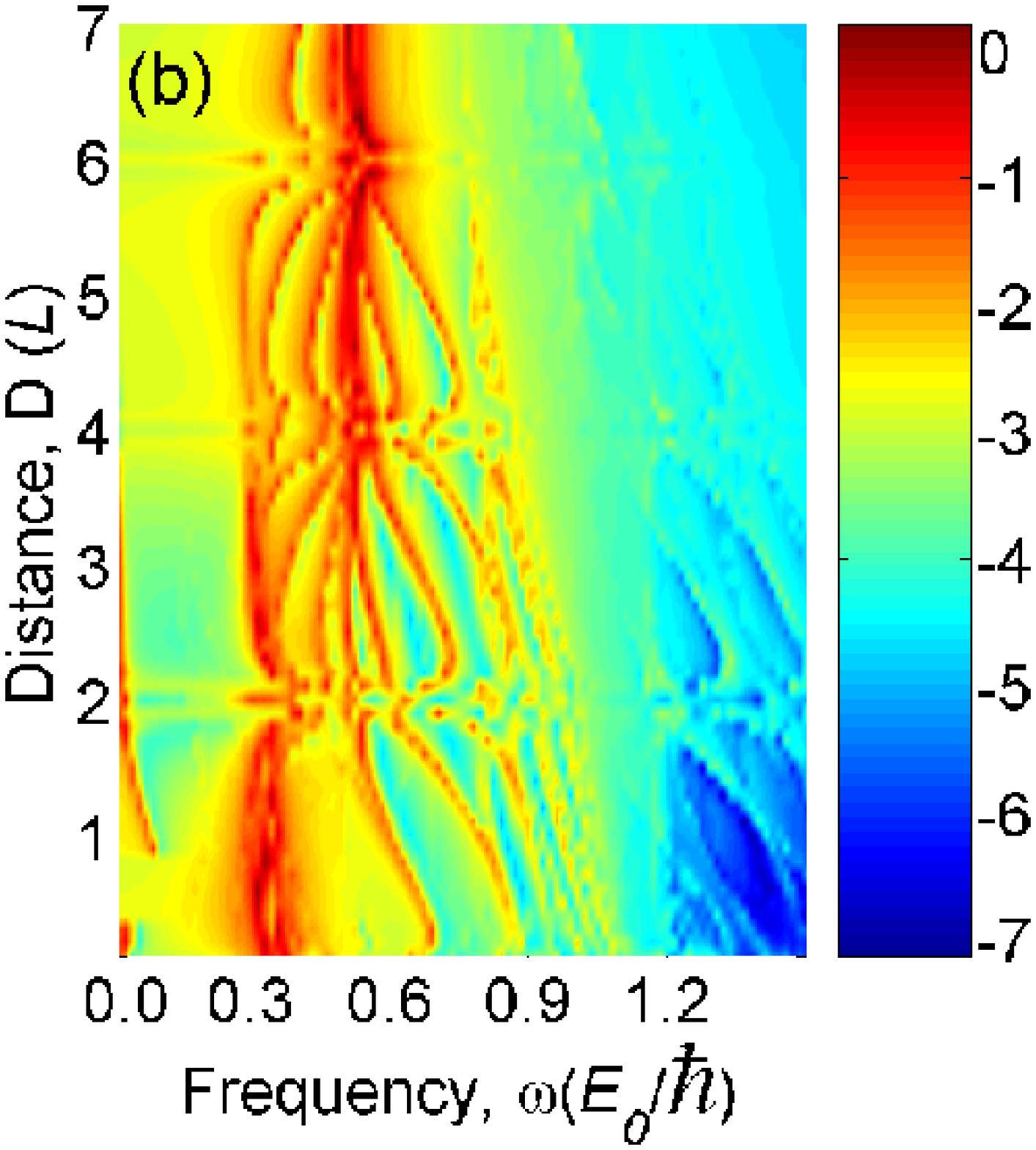} \vspace{%
-0.cm} \vspace{-0.5cm}
\caption{{\protect\footnotesize {\protect\vspace{-0.cm}Landscapes of the
induced electric field energy $\log_{10}(W_{\text{ind}})$ in systems of two
conducting nanospheres with radii $R=7L$ as a function of the frequency of
the external field $\protect\omega$ and the cluster separation distance $D$,
measured between the two closest points of the nanospheres. The direction of
the external field points along the line connecting the sphere centers,
damping constant $\protect\gamma=2\times 10^{-3} E_0$. (a) $N_{\text{el}}=20$%
, (b) $N_{\text{el}}=100$.}}}
\label{fig2}
\end{figure}

\textit{Dynamic Screening:} Although the characteristic Thomas-Fermi
screening length is known to increase with decreasing carrier
concentration, this quantity only describes the screening of slowly
varying potentials. The lower the carrier concentration, the worse
the system screens rapidly varying potentials. For relatively small
electron densities the screening length becomes comparable with the
distance between the spheres $D$. Thus, sensitivity of the response
of the system to carrier concentration suggests strong effects of
carrier screening. This will be most pronounced in the region
between the spheres, where the potential undergoes significant
changes. To illustrate how the carrier concentration in the
nanospheres dramatically changes the dynamic dielectric response of
the system, we show in Fig. 2 plots of $W_{\text{int}}$ as a
function of the frequency of the external field and relative
distance $D$ between the spheres. In Fig. 2(a) we consider the case
of the same low carrier concentration as in Fig. 1. There are
$N_{\text {el}}=20$ electrons in the system, with a characteristic
Fermi wavelength $\lambda_{\text F}\approx 3 L$, which is the same
order of magnitude as the radii of the spherical clusters. The
various observed resonances correspond to excitations of different
geometric modes available in the discrete spectrum of the system
(dipole-dipole, quadruple-quadruple, etc). For the parameters chosen
in Fig. 2(a), the dominant geometric resonances occur at frequencies
less than $\hbar \omega=0.2 E_0$. Interestingly, there is no
zero-frequency peak at $D\approx0$, which is in stark contrast to
the case of denser fillings that correspond to the classical limit,
e.g. as shown in Fig. 2(b). At low fillings, the delocalized charge
density response results in less efficient screening in the region
between the clusters, and hence the system of two clusters responds
as a whole. This significantly reduces the magnitude of the induced
charge densities near the closest surfaces of the spheres and limits
the maximum possible value of $W_{\text{int}}$. Moreover, quantum
discreteness of the energy levels results in a non-monotonic
dependence of $W_{\text{int}}$ on $D$, which in turn leads to the \textit{%
non-zero optimal distance} $D_{\text {\textrm{opt}}}$, discussed
above. Note, that for the parameter set chosen here at a finite frequency
$\omega=0.1 E_0$ the optimal distance is near $D=0$, i.e.
similar to the static response of the classical system.
In Fig. \ref{fig2}(b) we consider the same system parameters
but at a higher
carrier concentration, i.e. $N_{\text {el}}=100$ electrons. The
corresponding characteristic Fermi wavelength is $\lambda_{\text F}\approx 1
L$, which allows the dielectric response of the
system to be much closer to the classical limit. In this case many more
geometric resonances are observed compared to the low-filling regime
(Fig. \ref{fig2}(a)). Also, in contrast to the quantum limit these
resonances depend more strongly on changes in
separation $D$ and converge into a single dominant peak at $\hbar
\omega\approx 0.72%
$$
E_0$ for $D\ge 7%
$$
L$. Also note the large
maximum of $W_{\text{int}}$ at $D\approx0$ and $\omega\approx0$,
as it is expected for the static limit in the classical regime.

\textit{Optimal Design (static):} The above example illustrates that
there can be significant differences between the dielectric response
in the classical and quantum regimes. Let us now explore how the
quantum functionality of such structures can in principle be used
for the design of nanoscale devices. In the following, we pursue an
optimal design problem of a prototype system with multiple
adjustable parameters, using a numerical global optimization
technique based on the genetic algorithm \cite{thalken}.
Specifically, we wish to optimize a system containing $5$ point-like
charges with $q=+4e$ each. In order to reduce the complexity of the
problem the charges are placed on a line along the $z$ axis, and we
optimize the $z$ coordinates of the placed charges. This reduces the
optimization problem to $5$ parameters. A static external electric
field is applied along the $z$ axis. In order to discretize the
numerical search space, the positions of the charges are restricted
to be on a lattice with lattice constant $L$. We search for optimal
configurations of the charges that maximize the induced field
intensity in a target region $V_0$, located at the center of the
system at $z=1/2 L_{\mathrm{tot}}$. Here $L_{\mathrm{tot}}$ is the
total length of the optimization box along the $z$ direction. The
total number of electrons in the system $N_{\text{el}}=20$ is chosen
to insure the system's response to be in the quantum regime. It
usually takes about $1100$ seconds to perform optimization on a 20
node cluster configured with two 1GHz processors per node.

\begin{figure}[h]
\includegraphics[width=2.81cm,angle=0]{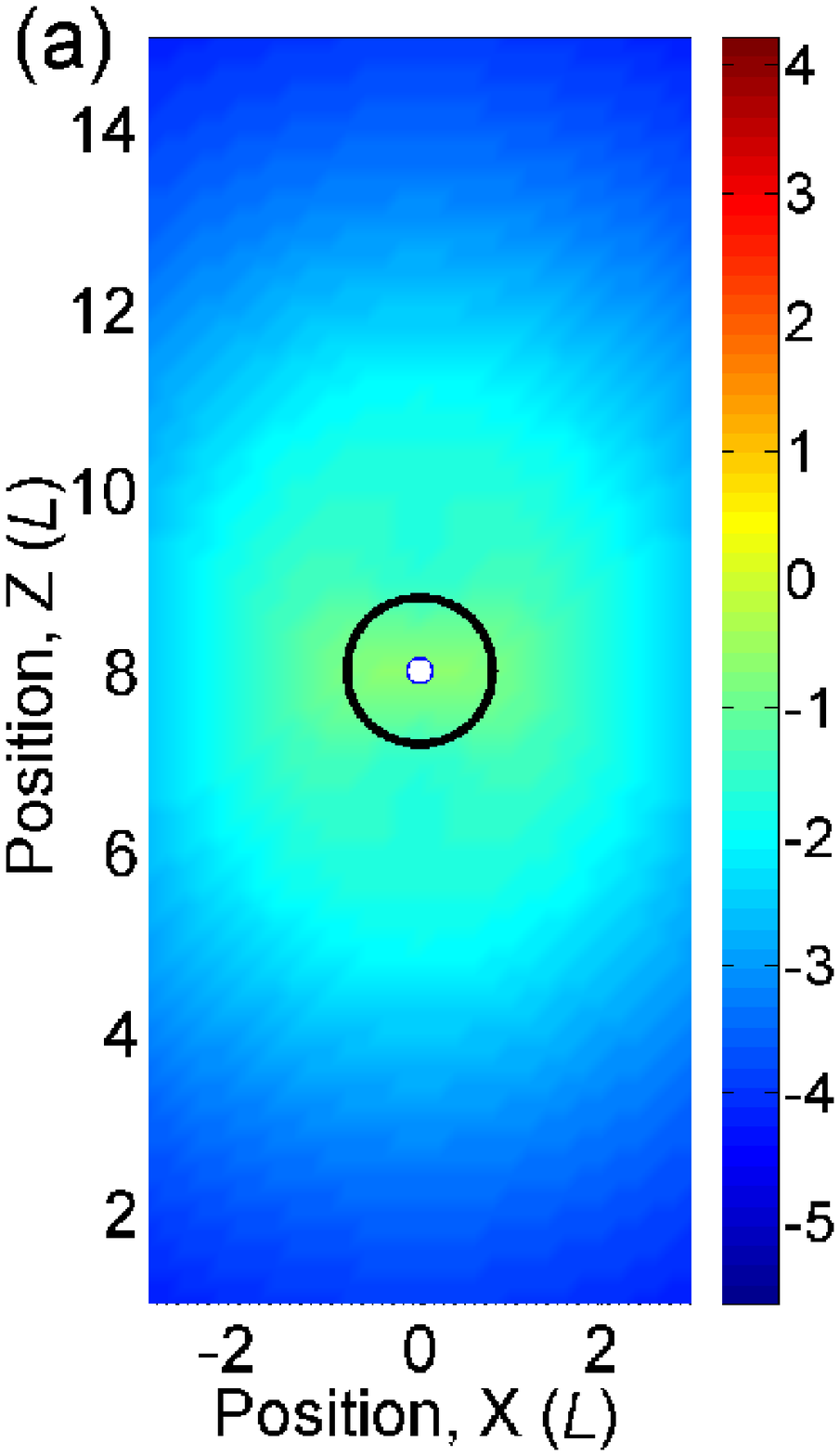} 
\includegraphics[width=2.81cm,angle=0]{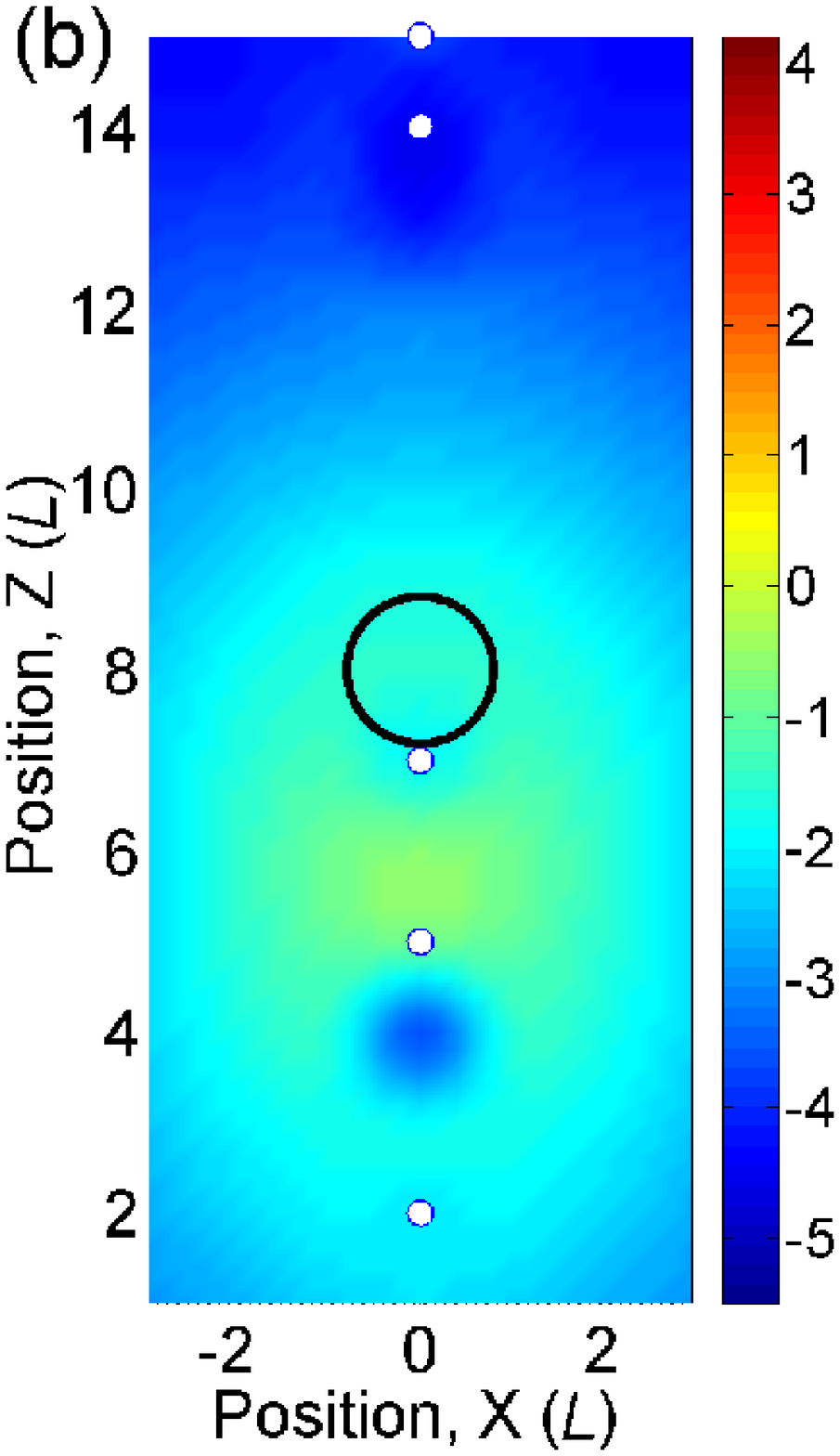}
\includegraphics[width=2.81cm,angle=0]{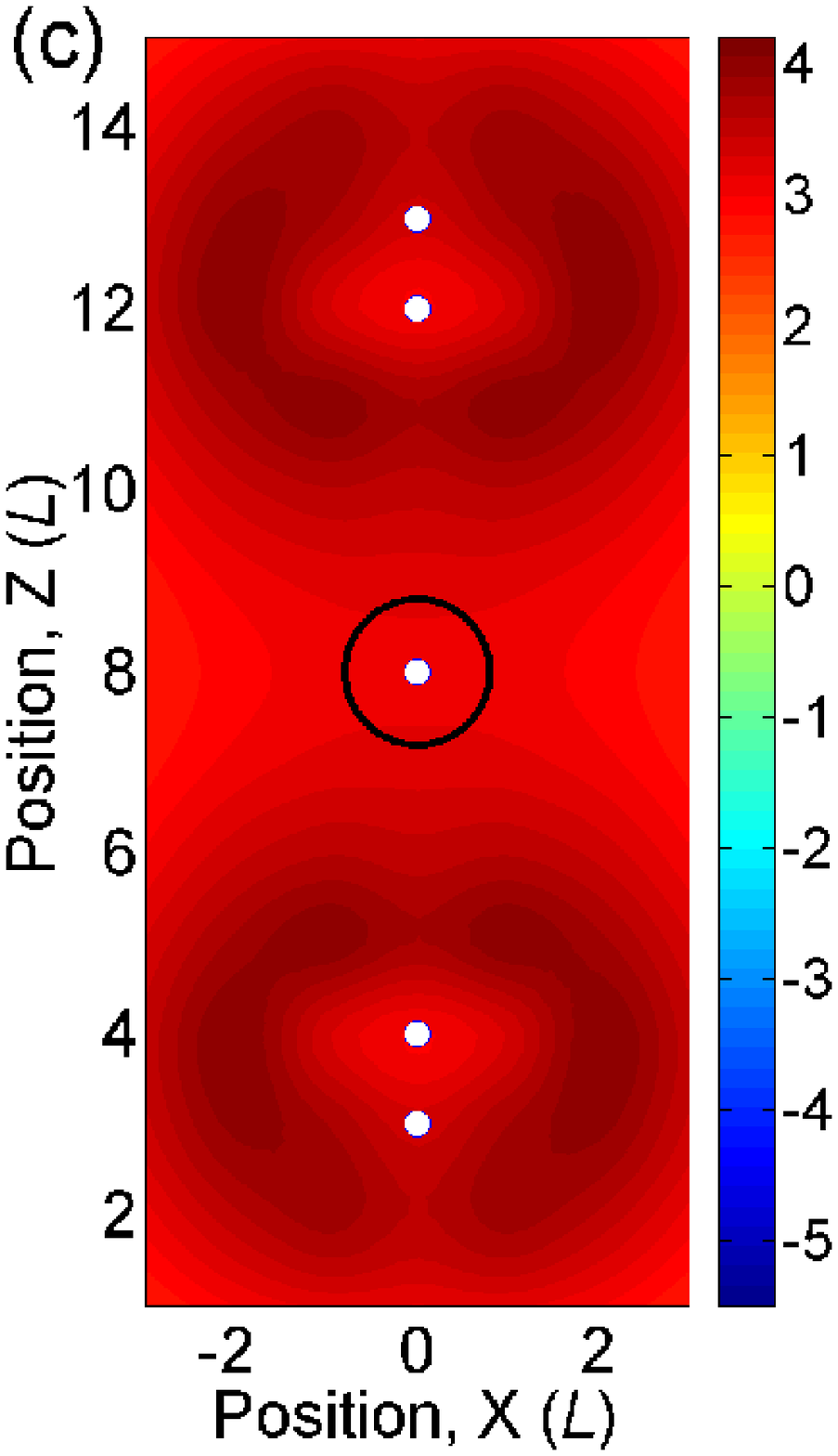}
\caption{{\protect\footnotesize {Induced field intensity $\log_{10}(|\mathbf{%
E}_{\text{ind}}(x,0,z)|^2)$ for different configurations of scatterers: five
point-like charges with $q=+4e$ (marked by white dots), each is placed along
a line on the z-axis, parallel to the external field. $N_{\text{el}}=20$ and
$\protect\gamma=2\times 10^{-3} E_0$. (a) All charges are placed at the
center, (b) a random configuration, (c) optimized configuration with target
to yield maximum intensity of induced field. }}}
\label{fig3}
\end{figure}

In Fig. 3(a) the intensity of the induced field is shown for the
case of all the point charges placed at the center of the target
area, as it would be suggested by classical intuition. While the
induced field is indeed largest at the system center, the overall
intensity is relatively small compared to optimized
configurations. For purposes of comparison, in Fig. 3(b) we also
show the intensity of the induced field for a random configuration
of point charges. In contrast to Figs. \ref{fig3}(a,b), Fig.
\ref{fig3}(c) displays the induced field for a numerically \textit{optimized
configuration} of point charges. In agreement with the
examples in Figs. 1 and 2(a), the optimal distance between the
placed charges is finite. The optimization algorithm finds a
compromise between the distance to the system center and the
inter-particle distances of the point charges that maximizes the
induced charge density. This is achieved via maximization of the
induced charge localization in the quantum system, leading to the
most efficient screening near the target volume. Thus, we find
that using a genetic algorithm it is possible to create highly
efficient optimized structures with broken spatial symmetries
which function as a sub-wavelength lens for electromagnetic
radiation. Note also that the optimal configuration in Fig. 3(c)
has an inversion symmetry about the system center which arises
naturally during the optimization procedure. Comparing the
optimized result with the classical configuration (Fig.
\ref{fig3}(a)) and the random configuration (Fig. \ref{fig3}(b)),
it is evident that the optimal configuration leads to a field
intensity in the focal region that is four orders of magnitude
larger. We have performed further optimizations for the positions
of 2-8 point-like charges with the same target functionality.
Significant even-odd effects are observed in the optimal
arrangements. For even numbers of charges, none of the charges
should be placed in the system center, whereas for odd numbers of
charges the optimal configuration consists of two symmetrically
arranged equal groups of charges, and one charge is placed in
center of the target area.

\begin{figure}[h]
\includegraphics[width=3.8cm,angle=0]{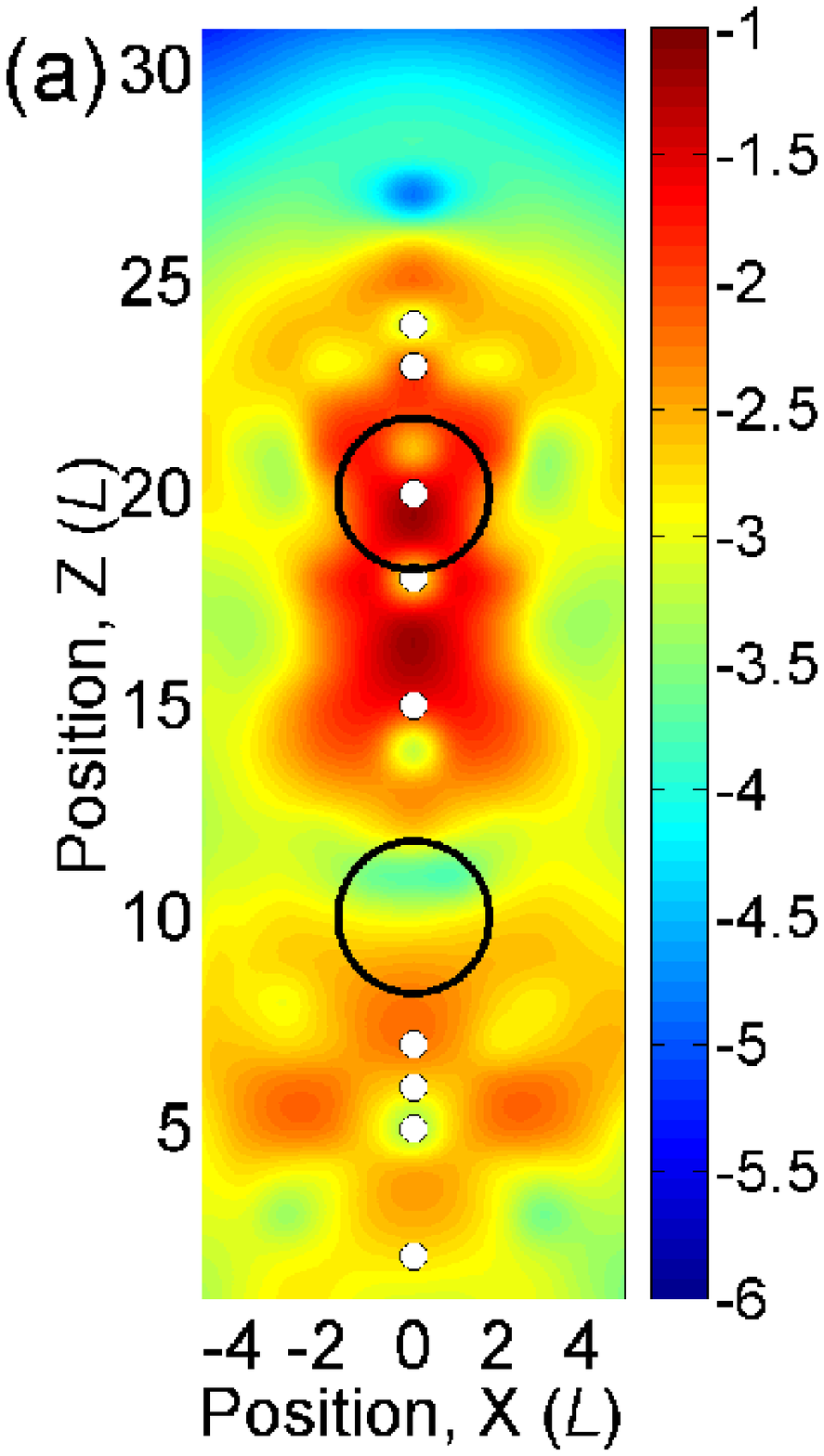}
\includegraphics[width=3.8cm,angle=0]{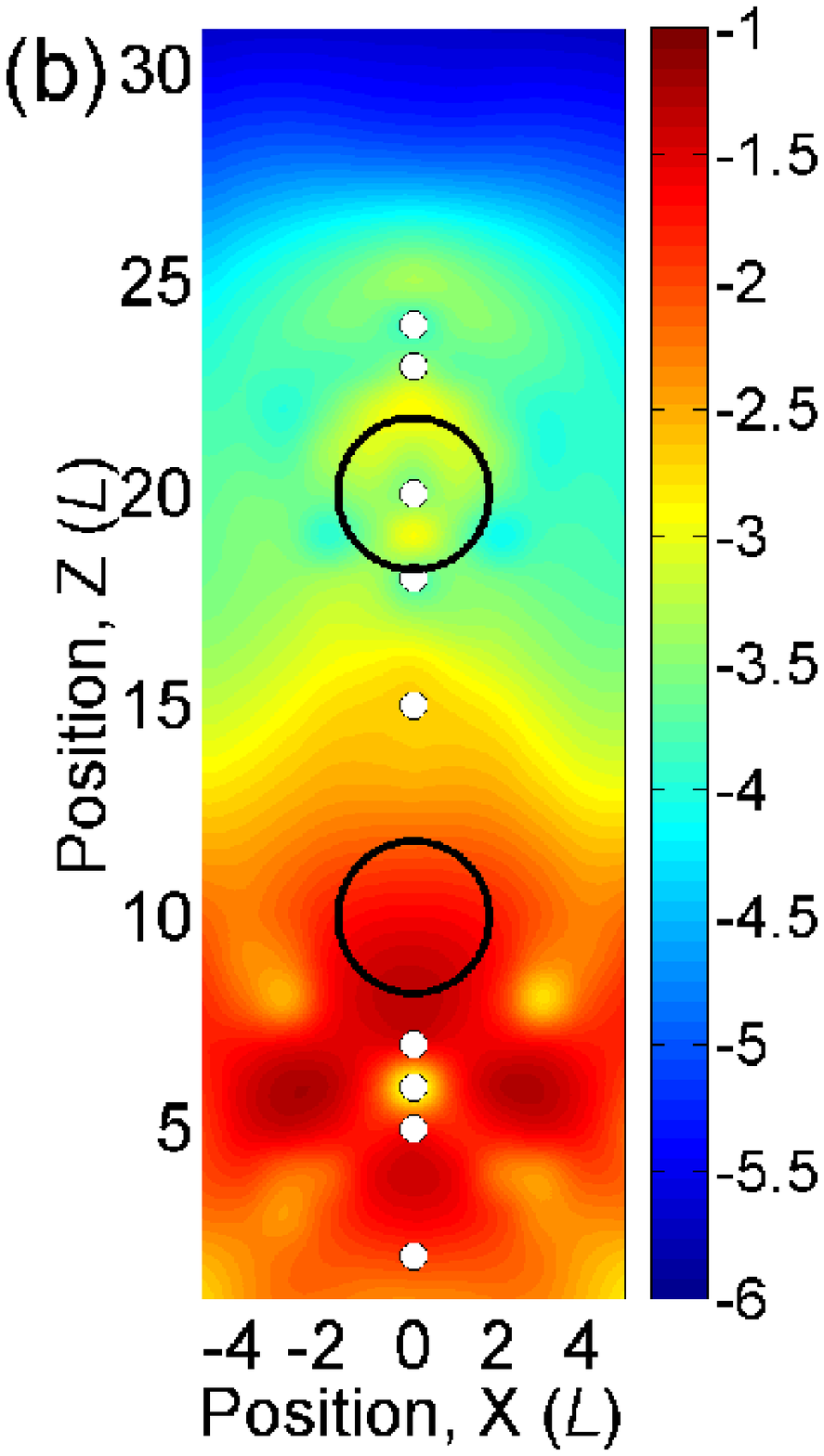} \caption{{\protect\footnotesize {\
``Frequency splitter": induced field intensities
$\log_{10}(|\mathbf{E}_{\text{ind}}(x,0,z)|^2)$ for optimized
configurations with $N_{\text{el}}=20$ electrons and $10$ moving
scattering centers (marked by white dots). Note that two moving
centers are overlapping at $z=5 L$. We use $\protect\gamma=2\times
10^{-3} E_0$, the
external field points along the $z$ direction. (a) $\hbar \protect\omega%
_1=0.09 E_0$, (b) $\hbar \protect\omega_2=0.130 E_0$. }}}
\label{fig4}
\end{figure}

\textit{Optimal Design (dynamic):} Finally, let us consider the case
of time-varying fields, with the goal to design a ``frequency
splitter" in the sub-wavelength limit. In this example we again
allow the point-like charges to be placed along the $z$ axis, and
search for optimal spatial configurations of the charges which
maximize the induced field intensity in a target volume centered
at $z=2/3 L_{\mathrm{tot}}$ for a field frequency $\omega_1$, and in
a second target volume centered at $z=1/3 L_{\mathrm{tot}}$ for a
second field frequency $\omega_2$. Numerical optimization was
performed for $10$ moving positive background charges with $q=+2e$
each and $N_{el}=20$ electrons in the system. In Fig. \ref{fig4},
we show the induced field intensity for optimized configurations
of charges with two different frequencies (a) $\hbar \omega_1=0.09
E_0$ and (b) $\hbar \omega_2=0.130 E_0$. The selectivity of this
device can be quantified by the induced field energies $W_1$ and
$W_2$ in the target volumes $1$ and $2$ correspondingly, and their
ratio at two distinct frequencies $\omega_1$ and $\omega_2$. We
find that for the optimized configuration the ratio
$W_1/W_2\approx 0.09$ at $\omega_1$, and $W_2/W_1\approx 0.07$ at
$\omega_2$.

\textit{Conclusions:}

In summary, the dielectric properties of spherical nanoclusters in
their quantum regime offer a richness of functionalities which is
absent in the classical case. These include a highly non-trivial
screening response and dependence on the frequency of the driving
field. Intuition based on classical field theory, e.g. divergence of
induced field in the static limit as the distance between the
spheres decreases, breaks down, and can thus not be relied on for
the design of new atomic-scaled devices. In particular, one cannot
expect to induce localized charge density distributions with a
characteristic length scale much smaller than the typical Fermi
wavelength of the system and collective excitation can be
dramatically modified. Moreover, in the quantum regime the
delocalized induced charge densities can provide \textit{increased
robustness} of the optimized quantity, and thus decrease the
complexity of optimal design. Quantum mechanical effects were also
found to set boundaries on the maximum values of target quantities,
i.e. induced field intensity in the system. Using genetic search
algorithms, we have demonstrated that optimal design can lead to
field intensities orders of magnitude larger than ``simple" guesses
based on intuition derived from classical theory.

For applications such as surface enhanced Raman scattering, the
electric field enhancement in the nm surrounding the molecule
depends critically on the local configuration of both the metallic
particle and the molecule. In principle, optimal particle shapes can
be found that maximize Raman activity for a given molecule attached
to the metal surface. The approach discussed in this paper opens new
possibilities for applications.  For example, it should be feasible
to use our theory in combination with search algorithms to create
robust and reliable highly Raman active nano-structured surfaces.

\textit{Acknowledgements:} We are grateful to Pinaki Sengupta for
fruitful discussions, and acknowledge support by the DOE under grant
DE-FG02-05ER46240. The simulations were carried out at the
high-performance computing computing center at USC. This work was
performed under the auspices of the National Nuclear Security
Administration of the U.S. Department of Energy at Los Alamos
National Laboratory under Contract No. DE-AC52-06NA25396 and
supported by DARPA and ONR.


\begin{thebibliography}{99}
\bibitem{footnote1} Consider a system of four charges: $+Q$ at positions
$z_1=-D/2-2R$, $z_3=D/2$, and $-Q$ at positions $z_2=-D/2$,
$z_4=D/2+2R$. The field intensity at
position $z$ ($D\gg R$) is $|E(z)|^2=16384 \:Q^2 R^2 z^2 (4 z^2+3
D^2)^2/(D^2-4 z^2)^6 +O(R^3)$, yielding $D^{-8}$ asymptotics.
\end{thebibliography}
\end{document}